\documentclass[conference]{IEEEtran}

\usepackage[utf8]{inputenc}

\ifCLASSINFOpdf
  \usepackage[pdftex]{graphicx}
  \graphicspath{{figures/}{./}}
  \DeclareGraphicsExtensions{.pdf,.jpeg,.png,.eps}
\else
  \usepackage[dvips]{graphicx}
  \graphicspath{{figures/}{./}}
  \DeclareGraphicsExtensions{.eps}
\fi

\usepackage[cmex10]{amsmath}
\usepackage{array}

\usepackage{mdwmath}
\usepackage{mdwtab}
\usepackage{fixltx2e}

\usepackage{stfloats}

\usepackage{url}

\usepackage{epstopdf}
\usepackage{booktabs}

\usepackage{hyperref}

\begin{document}
\title{Air-coupled Thickness Measurements\\ of Stainless Steel}

\author{\IEEEauthorblockN{Grunde Waag, Lars Hoff}
\IEEEauthorblockA{Institute of Micro and Nano Systems Technology\\
Vestfold University College\\
Horten, Norway\\
grunde.waag@hive.no}
\and
\IEEEauthorblockN{Petter Norli}
\IEEEauthorblockA{Det Norske Veritas\\
Høvik, Norway}}

\maketitle
\begin{abstract}
A method of measuring the thickness of steel plates using through transmission of an acoustic pulse is demonstrated. This study has been done on a stainless steel plate with regions of thickness $10.0~\rm mm$, $9.8~\rm mm$, and $9.6~\rm mm$, using broadband pulses with energy in $200~\rm kHz$ to $600~\rm kHz$ band. Ultimately the goal is to perform similar air-coupled thickness measurements in a single sided pitch-catch measurement setup.

The spectra of the transmitted pulses show the first and second harmonics of the compressional waves in the plate. When compared to a plane wave model of a fluid layer embedded in air, the second harmonic of the plate resonance fits well with the expected value. However, the first harmonic deviates such that the plate appears thicker at this resonance. This is believed to be caused by the finite aperture of the transmitting transducer, causing deviations from a plane wave. Thickness differences of $0.2~\rm mm$ between the different regions of the plate were shown to be resolved.

A third peak was found in the spectra. The origin of this peaks has not been verified, but is believed to come from the third harmonic of the shear wave in the steel plate.
\end{abstract}

\section{Introduction}
Ultrasound inspection techniques have been used in the non-destructive test and evaluation community for decades. Traditionally ultrasonic inspection techniques have used liquid couplants to overcome the impedance barrier between air and steel. However, because of the drawbacks associated with liquid couplants, i.e. preparation time, loss of portability, and the couplants effect on the target, the NDT community has been motivated to investigate the potential of air-coupled ultrasound techniques \cite{grandia_nde_1995}. Air-coupled ultrasound enables quick, non-contact measurements at a distance. It has been successfully applied for inspection of several materials, such as aluminum, epoxy, food and textiles \cite{gomez_alvarez-arenas_simultaneous_2010, grandia_nde_1995, alvarez-arenas_non-contact_2009, blomme_air-coupled_2002}.

A broadband acoustic pulse transmitted into a plate will be filtered by the frequency response of the transmission coefficient, giving resonance peaks at frequencies where the plate thickness is an integer number of half-wavelengths. This is a well-known principle in sound propagation in layered media. The distance between the peaks in the spectrum of the plate response can be used to calculate the thickness when the speed of sound in the material is known, and vice versa.

Det Norske Veritas have employed this principle for inspection of steel pipes, using water or high pressure natural gas as coupling medium. The goal of this paper is to apply this principle to measure the thickness of steel plates with air as the coupling medium. The transmission loss from air to steel is extremely high, approximately $-45~\rm dB$, giving challenges in getting sufficient energy into the steel, and causing strong reverberations that interfere with the weak resonance from the plate.

\section{Theory}
The transmission coefficient for a fluid layer embedded in a fluid \cite{brekhovskikh_acoustics_1998} is given by:

\begin{equation}
W = \frac{4 Z Z_{1}}{(Z - Z_{1})^2\exp(i\varphi) + (Z + Z_{1})^2\exp(-i\varphi)},
\label{eq_W}
\end{equation}

where the impedance of the fluid layer is $Z_1 = \frac{\rho_1c_1}{\cos\theta_1}$, the impedance of the surrounding fluid is $Z = \frac{\rho c}{\cos\theta}$, and $\varphi = k_1 d \cos \theta_1$. $\theta$ is the angle of incidence and $\theta_1$ is the propagation angle in the intermediate layer. When $\varphi = k_1 d \cos \theta_1 = n\pi$, and $n = [1, 2, 3,...]$, there is a resonance. At normal incidence, $\theta = 0$, these resonances occur at frequencies, $f_n$, given by:

\begin{equation}
f_n = \frac{n c}{2 d}.
\label{eq_thickness}
\end{equation}

\begin{figure}[!h]
\centering
\includegraphics[width=2.5in]{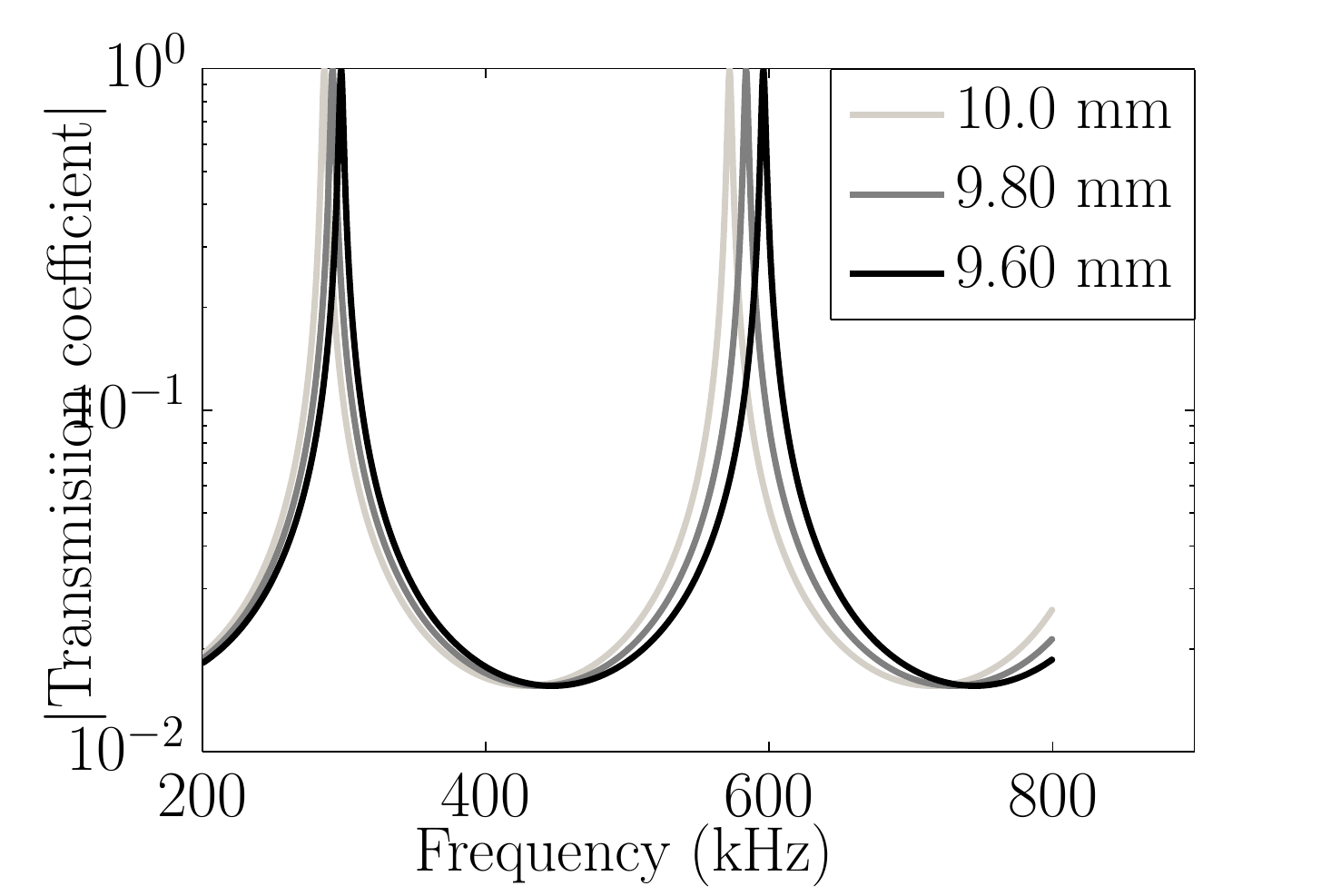}
\caption{Transmission coefficients for a $10~\rm mm$, $9.8~\rm mm$, and $9.6~\rm mm$ thick plate embedded in air for plane waves at normal incidence.}
\label{fig_transmission}
\end{figure}

When sending a pulse towards the plate and receiving the through transmitted pulse, peaks will be found in the transmission spectrum at the resonance frequencies of the plate. Between these frequencies the transmission coefficient will be close to zero, and little sound energy is transmitted, see Fig. \ref{fig_transmission}. If the speed of sound of the material is known the thickness can be computed from the positions of the peaks in the transmission spectrum. If the thicknesses is known, the speed of sound can be computed.

\section{Experiments}
\begin{figure}[!h]
\centering
\includegraphics[width=2.5in]{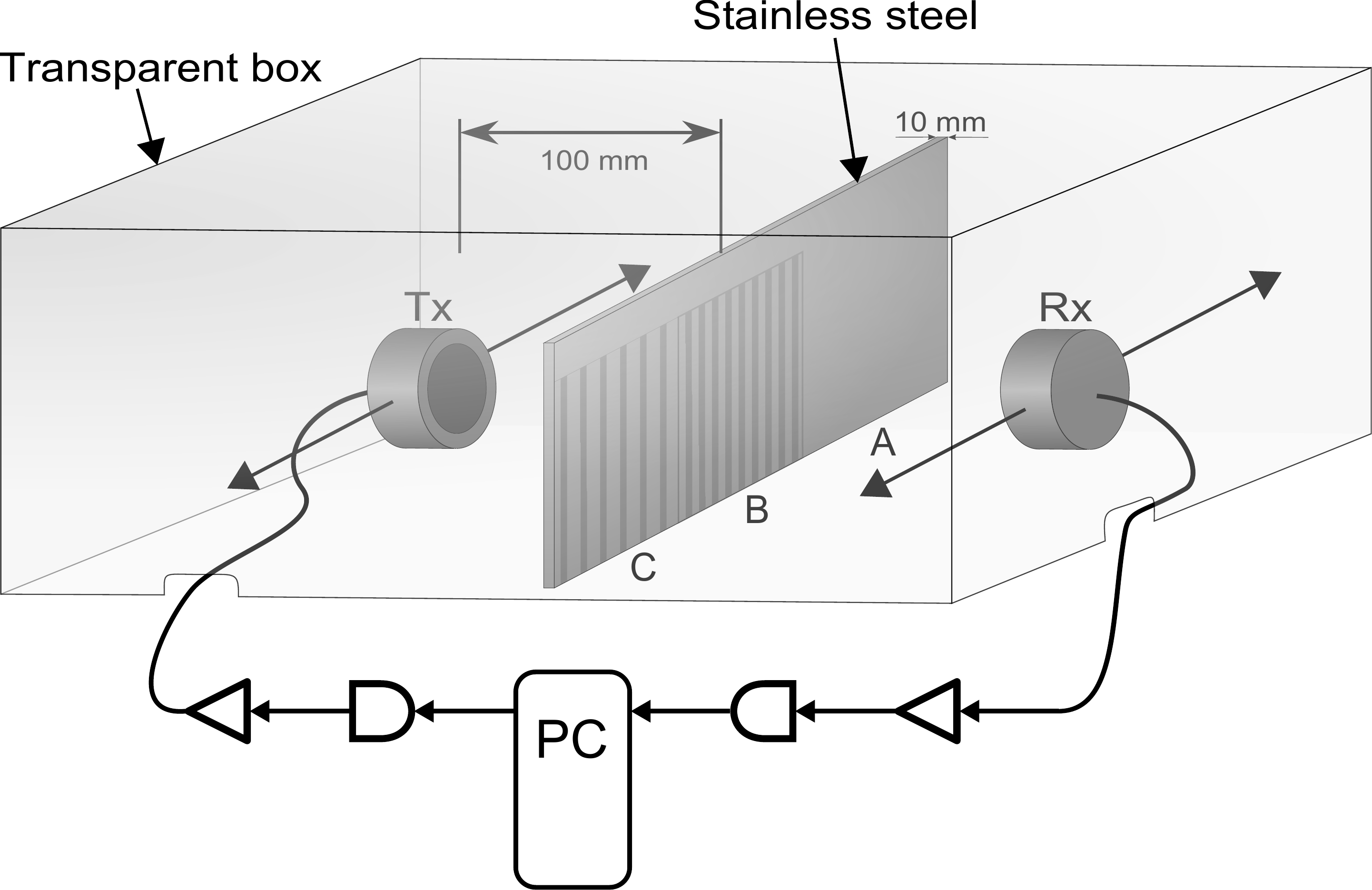}

\caption{Experimental setup. Regions A, B, and C are $10~\rm mm$, $9.8~\rm mm$ and $9.6~\rm mm$ thick respectively. The distance from the transducers to target is approximately $100~\rm mm$.}
\label{fig_experiment}
\end{figure}

The experimental setup is illustrated in Fig. \ref{fig_experiment}. A transducer is mounted on either side of a stainless steel plate. A broad band pulse is sent from one side, transmitted through the plate, and is received and recorded on the opposite side.

One transducer is used for transmitting pulses (Tx). This transducer is custom built for use in high pressure gas. It consists of two elements, one center disc and one annulus, but only the center disc is used in these experiments. The center frequency of the transducer is around $540~\rm kHz$ and the diameter of the center disc is $18~\rm mm$.

The receiving transducer (Rx) is a NCT500-D6 produced by Ultran Group (The Ultran Group Ltd., State College, PA, USA), with center frequency at $500~\rm kHz$. The diameter of the active element is $6~\rm mm$. Both transducers have plane elements without focus.

The Rx transducer was connected to a preamplifier, Olympus Panametrics 5662 (Olympus NDT Inc., Waltham, MA, USA) with $54~\rm dB$ gain and to a NI-PXI 5922 AD-converter (National Instruments Inc., Austin, TX, USA).

A waveform generator, NI-PXI 5421 (National Instruments Inc., Austin, TX, USA), and a E\&I 2100L (Electronics \& Innovation, Rochester, NY, USA) power amplifier were used. The transducer was connected through an electric matching network to the power amplifier.

The transducers were moved parallel to the steel plate, but always so that they were fixed along the same acoustical axis, which is normal to the steel plate. The distance between the transducers and the steel plate on both sides of the plate was held fixed to $100~\rm mm$.

A transparent box was placed on top of the setup to minimize the convection of the air between the transducers and the steel plate.


The target was a steel plate with nominal thickness of $10.0~\rm mm$. Two rectangular areas were machined down $0.2$ and $0.4~\rm mm$. Hence the plate consists of three areas, A, B and C, with thickness $10.0~\rm mm$, $9.8~\rm mm$ and $9.6~\rm mm$ respectively. By moving the transducers parallel to the steel plate, the three different areas were inspected.

Measurements were done by sending pulses from the transmitter through the steel target to the receiver. Linear chirps with frequencies from $200$ to $600~\rm kHz$ were used. 200 recordings were averaged to improve signal-to-noise ratio. The averaged signal was time-gated by a window starting just before the arrival of the transmitted pulse. The window had a duration $130~\rm ms$. The time-gating of the signal was done to isolate the transmitted pulse, reducing noise from reflections and electromagnetic coupling between transmitter and receiver. The power spectrum was estimated from this time window using a periodogram with a Hanning window. Measurements were done in each of the three regions A, B and C in  Fig. \ref{fig_experiment}, by manually moving the transducers parallel to the steel plate.

\section{Results}

\begin{figure}[!b]
\centering
\includegraphics[width=2.5in]{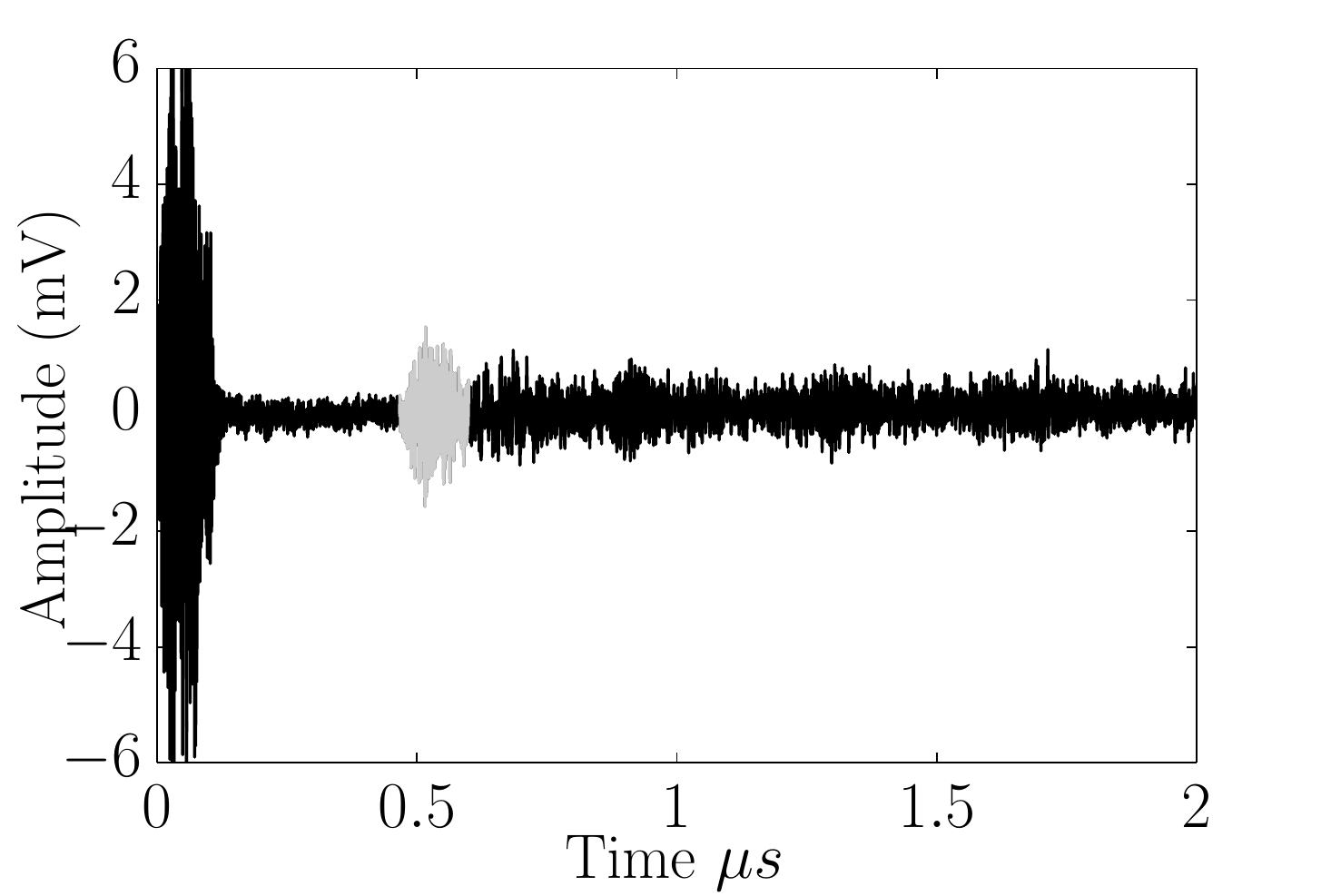}
\caption{Time signal for a measurement in the region A. The frequency spectrum was estimated for the part of the pulse shown in gray. The first spike is electromagnetic interference. The through transmitted signal arrives at approximately $0.57~\rm ms$.}
\label{fig_timesignal}
\end{figure}

A typical time series is shown in Fig. \ref{fig_timesignal} for a measurement in region A ($10~\rm mm$). The window used when calculating the spectrum is marked on Fig. \ref{fig_timesignal}. The strong signals at the start of the signal, at approximately $0.2~\rm ms$, is caused by electromagnetic coupling between the transmitter and receiver.

\begin{figure*}
\centering
\includegraphics[width=4.1in,angle=-90]{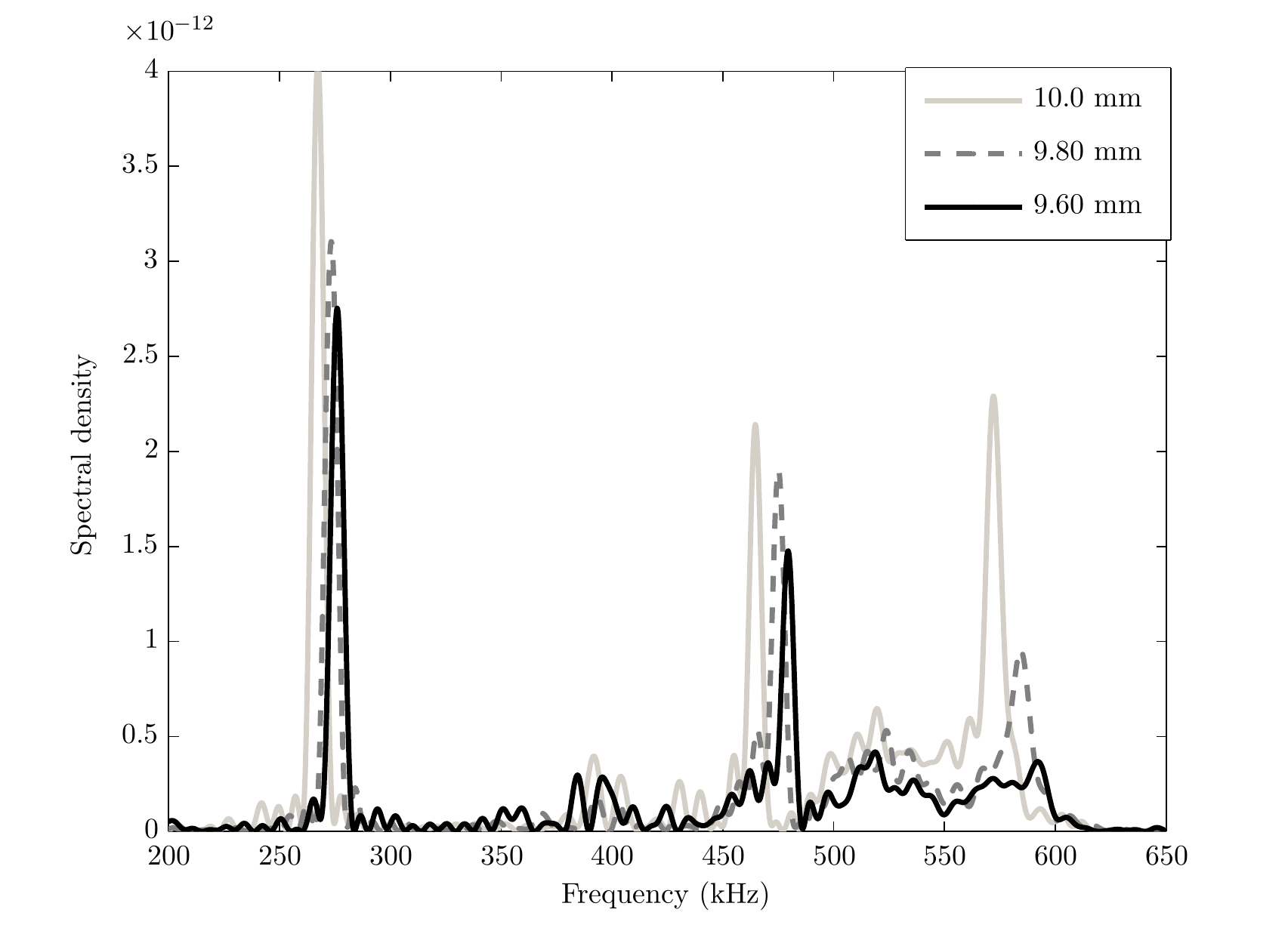}
\caption{Transmission spectrum of the signals for $10~\rm mm$, $9.8~\rm mm$, and $9.6~\rm mm$. The first harmonic is observed just under $300~\rm kHz$ and the second harmonic is observed at just below $600~\rm kHz$. A third set of peaks are observed at around $460~\rm kHz$.}
\label{fig_tailspectrum}
\end{figure*}

The spectrum of the time gated signal is shown in Fig. \ref{fig_tailspectrum} for the measurements on region A, B and C ($10~\rm mm$, $9.8~\rm mm$, and $9.6~\rm mm$ thickness).

Three peaks are observed in the transmission spectrum, Fig. \ref{fig_tailspectrum}. The first (approx. $270~\rm kHz$) and the last (approx. $570~\rm kHz$) observed peaks are identified as the first and the second harmonic resonance frequency respectively. The second peak does not match any resonance frequency expected from the fluid layer model. It is assumed to be the third harmonic of the shear waves.

The speed of sound in the plate was estimated from the second harmonic peak from a measurement in region A, where the thickness is assumed to be known to be $10.0~\rm mm$. Using (\ref{eq_thickness}) the speed of sound was calculated to be $c = 5720~\rm ms^{-1}$.

Thicknesses were then calculated for each of the peaks in each of the transmission spectra according to (\ref{eq_thickness}) using the speed of sound found above. The measured peak frequencies and computed thicknesses for area A, B and C are summarized in Table \ref{table_result}.

\section{Discussion}
From Table \ref{table_result} the peaks in the tail spectrum are observed to match well with the thickness for the second harmonic. The deviation from the nominal thickness is approximately $1\%$. The first harmonic gives thicknesses that are approximately $7\%$ higher than the nominal thicknesses. This might be explained by the fact that the fluid model, (\ref{eq_thickness}), assumes plane waves with normal incidence. Because of the finite size of the source, the acoustic field will not be perfectly plane, but has an opening angle given by $\sin\nu_{3\rm{dB}} = 0.51 \lambda D^{-1}$. $D$ is the active diameter of the transducer. At $250~\rm kHz$ this gives $\nu_{3\rm{dB}} = 2.3^o$. Hence at oblique angle of incidence the peaks will be shifted down. Improving the model according to this, it should be possible to increase the accuracy of the thickness estimate.

The second peak observed for each region cannot be explained by the simple fluid model usid in (\ref{eq_thickness}). However, it might be explained by including shear waves in the steel plate, $c_T \approx 0.55c = 3146~\rm ms^{-1}$, and $n = 3$ in (\ref{eq_thickness}). This gives the thicknesses: $10.16~\rm mm$, $9.93~\rm mm$ and $9.84~\rm mm$, see Table \ref{table_result}. Further investigations are needed to confirm this, but if confirmed, this can also be used to improve the thickness estimates.

The resolution of the setup can clearly distinguish between steel plates where the thickness difference is $2\%$ or $0.2~\rm mm$. The absolute accuracy of the thickness measurements depends on the accuracy of the speed of sound in the plate and estimation of the resonance frequencies.

\begin{table}[!h]
\centering
\caption{Extracted peaks from the tail spectrum. Note that the 3rd peak, Region A is used to estimate the speed of sound and is not a measurement of the thickness.}
\label{table_result}
\begin{tabular}{@{} r r r @{}}
\toprule
\multicolumn{3}{c}{Region A ($10.0~\rm mm$)}\\
& Frequency ($\rm kHz$) & Thickness ($\rm mm$)\\ \midrule
1st Peak & 267.33 & 10.70\\
2nd Peak & 464.63 & 10.16\\
3rd Peak & 571.98 & 10.0\\ \midrule
\multicolumn{3}{c}{Region B ($9.8~\rm mm$)}\\
& Frequency ($\rm kHz$) & Thickness ($\rm mm$)\\ \midrule
1st Peak & 273.51 & 10.46\\
2nd Peak & 475.16 &  9.93\\
3rd Peak & 584.56 &  9.79\\ \midrule
\multicolumn{3}{c}{Region C ($9.6~\rm mm$)}\\
& Frequency ($\rm kHz$) & Thickness ($\rm mm$)\\ \midrule
1st Peak & 276.03 & 10.36\\
2nd Peak & 479.51 & 9.84\\
3rd Peak & 592.0  & 9.66\\
\bottomrule
\end{tabular}
\end{table}

\section{Conclusion}
We have demonstrated a method to measure the thickness of steel plates in air, using air-coupled ultrasound with through transmission.

The method can clearly distinguish between absolute difference in thickness of $0.2~\rm mm$, $2\%$ relative.

Improvements to the accuracy of this method can be acheived by including the effects of shear waves and a finite source in the model.

\bibliographystyle{IEEEtran}
\bibliography{IEEEabrv,biblio}

\end{document}